\documentstyle[aas2pp4]{article}

\slugcomment{}

\lefthead{Wada, Spaans and Kim}
\righthead{Multi-phase gas dynamics}

\newcommand{\beq}{\begin{equation} }
\newcommand{\eeq}{\end{equation}}
\newcommand{\bea}{\begin{eqnarray} }
\newcommand{\eea}{\end{eqnarray}}
\def\mbf#1{\mbox{\boldmath ${#1}$}}
\newcommand  \hi    {H\,{\sc i}}

\begin{document}
\title{Formation of Cavities, Filaments, and Clumps by
the Non-linear Development of Thermal and Gravitational
Instabilities 
in the Interstellar Medium under Stellar Feedback}
\author{Keiichi Wada\altaffilmark{1}, Marco Spaans\altaffilmark{2}, and
Sungeun Kim\altaffilmark{3}}
\affil{$^1$National Astronomical Observatory, Mitaka, 181-8588, Japan\\
Email: wada.keiichi@nao.ac.jp}
\affil{$^2$Harvard-Smithsonian Center for Astrophysics, 60 Garden Street, Cambridge, MA 02138}
\affil{$^3$Astronomy Department, University of Illinois at
Urbana-Champaign, Urbana, IL 61801}




\begin{abstract}
Based on our high resolution, two-dimensional hydrodynamical
simulations, we propose that large cavities may be formed by the nonlinear development of
the combined thermal and gravitational instabilities, without need for
stellar energy injection in a galaxy modeling the Large Magellanic Clouds (LMC).
Our numerical model of the star formation allows us to 
follow the evolution of the blastwaves due to supernovae in 
the inhomogeous, multi-phase, and turbulent-like media self-consistently.
Formation of kpc-scale inhomogeneity, such as cavities, observed \ion{H}{1} map of the LMC, is suppressed by 
frequent supernovae (average supernova rate for the whole disk 
is $\sim 0.001$ yr$^{-1}$). However the supernova explosions are necessary for
the hot component ($T_g > 10^{6-7}$ K).
Position-velocity maps show that kpc-scale shells/arcs formed 
through the nonlinear evolution in a model without stellar 
energy feedback has similar kinematics to explosional phenomena, such as
supernovae.
We also find that dense clumps and filamentary structure are 
formed due to a natural consequence of the non-linear evolution of the
multi-phage ISM. Although the ISM in a small scale looks turbulent-like and
transient, the global structure of the ISM is quasi-stable. In the quasi-stable
phase, the volume filling factor of the hot, warm, cold components are
$\sim 0.2, \sim 0.6$, and  $\sim 0.2$, respectively.
We compare the observations of \ion{H}{1} and molecular gas of the 
LMC with the numerically obtained \ion{H}{1} and CO brightness temperature distribution.
The morphology and statistical properties 
of the numerical \ion{H}{1} and CO maps are discussed.
We find that the cloud mass spectrum of our models
represent a power-law shape, but their slopes change between models with and 
without the stellar energy injection, and also the slope depends on the
threshold brightness temperature of CO.

\end{abstract}


\keywords{ISM: structure, kinematics and dynamics --- galaxies:
structure --- individual: LMC --- method: numerical}

\section{INTRODUCTION}

The topology of the neutral 
interstellar medium (ISM) can be studied in great detail 
by the spatial and velocity structures in the neutral \ion{H}{1} gas.
A recent 
high-resolution \ion{H}{1} survey of the Large Magellanic Cloud (LMC) reveals  
that the structure of the neutral atomic inter stellar gas is dominated by 
numerous holes and shells as well as complex filamentary structure (\cite{KIM98}). These features are commonly seen in recent high-resolution \ion{H}{1} images of 
nearby galaxies obtained with radio synthesis interferometers
(\cite{DD90}; \cite{P92}; \cite{SS97}; \cite{WB99}; \cite{ST99}). 
In general, the shell-like and hole structure seen in
\ion{H}{1} has been understood as the cumulative effect of 
stellar winds from massive stars and supernova explosions evacuating the cool 
ISM (\cite{TT88}; \cite{VDH96}; \cite{OEY96} ; \cite{OC97}). However, the extensive study of \ion{H}{1} shells in the LMC shows that 
there is a relatively weak correlation between the \ion{H}{1} shells and the 
ionized gas traced out by the \ion{H}{2} regions and \ion{H}{2} filaments 
(\cite{KI99}). Furthermore, the correlation between the \ion{H}{1} 
shells and 122 OB stellar associations in the LMC is not very tight (\cite{KI00}). Rhode et al. (1999) claimed that there is no remnant 
star clusters at the center of the \ion{H}{1} holes in Holmberg II, 
and it is inconsistent with the SNe hypothesis.
Moreover an energy source generating 
the kpc-scale supergiant \ion{H}{1} holes is a puzzle.
These issues raise an interesting question about whether \ion{H}{1} 
shells/holes  have been formed by the interaction between stars and the ISM or not.

Recent hydrodynamical simulations by Wada \& Norman (1999) demonstrate that 
a gravitationally and thermally unstable disk, which models an ISM in
galaxies,  can generate the cold, dense 
clumps and filaments surrounded by hot, diffuse medium. 
They show that porous structure is a natural consequence
of the non-linear evolution of the ISM.
This result strongly 
suggests that some fraction of \ion{H}{1} shells, supershells or
holes seen in galaxies does not relate to the  
interaction between stellar activities and the ISM.

The other important component of the ISM is the molecular gas.
Molecular clouds are potential sites of star formation,
but their formation mechanism, and the relationship between their evolution
and star formation have been poorly understood.
Dense molecular hydrogen  is traced by the rotational transition of 
CO, and a recent high resolution survey of $^{12}$CO (J$=$1-0) in 
the LMC with NANTEN, which is 
4m millimeter-wave telescope at Las Campanas Observatory, established 
a comprehensive view of the giant molecular clouds in the LMC (\cite{FU98}). Since the gas clumps, whose size is typically $10-100$ pc,
 in the simulations by Wada \& Norman (1999)
are dense ($n > 1000$ cm$^{-3}$) and cold ($T=10-100 $ K), they are
counterparts of the observed giant molecular clouds in the LMC or other galaxies.

Using numerical simulations,
 Feitzinger et al. (1981) and Gardiner, Turfus, \& Putman (1998)
have studied global dynamics of the ISM and star formation in
the LMC. 
Although the numerical
methods they used are different, the stochastic self-propagating
star formation model (e.g. Seiden \& Gerola 1984) in  Feitzinger et al. (1981) and 
sticky particle method in Gardiner, Turfus, \& Putman (1998),
are both phenomenological concerning	
the structure and dynamics of the
ISM and star formation processes. 
Unfortunately the spatial resolution (100 pc in Feitzinger et al. 1981) 
and  mass resolution (6$\times 10^4 M_\odot$, which corresponds 
to $\sim 200$ pc for $n\sim 1$ cm$^{-3}$ and $H \sim 100$ pc, in Gardiner, Turfus,
\& Putman 1998) 
in these simulations are not good enough for comparison between
models and the recent high resolution observations ($\sim$ 15 pc for HI
(\cite{KIM98}) and $\sim $ 40 pc for CO (J$=$1-0) (\cite{FU98})).

In this paper, we apply the numerical scheme used in Wada \& Norman
(1999) to a LMC-type model galaxy, and we  
conduct two-dimensional hydrodynamical simulations of 
the multi-phase ISM in a LMC-like model galaxy, 
taking into account self-gravity of the gas, radiative cooling and
various heating processes, such as supernova explosions.
High spatial resolution (7.8 pc) and 
a modern hydrodynamical scheme allow us to model the star formation and 
its feedback less phenomenologically, and therefore it needs less assumptions
than the previous semi-analytic and numerical approaches (e.g. a review
by \cite{SF95}).
In contrast to the model used in Wada \& Norman (1999), 
the rotation curve assumed here is nearly rigid as suggested in the LMC
or other LMC-type dwarf galaxies.
We derive the \ion{H}{1} and CO brightness map from 
the simulations. Then we compare these simulation results with the \ion{H}{1} 
and CO observations of the LMC.

In \S 2, we describe our numerical method and
models. In \S 3, the numerical results with and without star formation
are discussed on morphology and statistical structure of the ISM, then they are compared with the observations.
Position-Velocity diagrams derived from the numerical results
are also discussed.
Comparison with past numerical simulations and 
other implications are discussed in 
\S 4, and conclusions are presented in \S 5.
\section{NUMERICAL METHOD AND MODELS}
Taking into account the multi-phase and inhomogenous nature of the ISM is
crucial for realistic simulations of global star formation in galaxies.
Stars are formed preferentially in cold molecular gas,
 and supernovae produce 
low density, high temperature gas. Interaction between such different 
phases are also impornant (e.g. \cite{MO}; \cite{IH}).
In order to simulate dynamics of the inhomogeous, multi-phase
interstellar matter and star formation, our numerical method has
following features.
(1) High spatial resolution (7.8 pc), for diffuse
gas to high density gas. We use an Euler mesh code with 
1024$^2$ Cartesian grid points.
(2) The simulations are global in order to study effects of galactic
rotation and other kpc scale phenomena to the local structure of the ISM.
(3) Self-gravity of the gas is calculated. 
(4) Radiative cooling for gas whose temperature is between $10^8$ K and
10 K is taken into account.
(5) Various heating processes are included. Here we assume photoelectric heating due to dust grains and UV radiation, SN explosions, and stellar wind from massive stars. 
(6) High numerical accuracy for shocks is achieved with
a modern hydrodynamical scheme. This is crucial,
because the ISM is usually supersonic, and SNe produce
strong shocks.

\setcounter{footnote}{0} We solve the following hydrodynamical 
equations and the Poisson equation numerically
in two dimensions to simulate the evolution of a rotating gas disk in 
a stellar potential.

\begin{equation}
\frac{\partial \rho}{\partial t} + \nabla \cdot (\rho \mbf{v}) = 0,
\label{eqn: rho} 
\end{equation}

\begin{equation}
\frac{\partial \mbf{v}}{\partial t} + (\mbf{v}
\cdot \nabla)\mbf{v} +\frac{\nabla p}{\rho} + \nabla \Phi_{\rm ext} +
\nabla \Phi_{\rm sg} = 0, \label{eqn: rhov}
\end{equation}

\begin{equation}
 \frac{\partial
E}{\partial t} + \frac{1}{\rho} \nabla \cdot [(\rho E+p)\mbf{v}] =
\Gamma_{\rm UV} + \Gamma_{\rm \star}-\rho \Lambda(T_g), \label{eqn: en}
\end{equation}

\begin{equation}
 \nabla^2 \Phi_{\rm sg} = 4 \pi G \rho, \label{eqn: poi} 
\end{equation}

 where, $\rho,p,\mbf{v}$ are the density, pressure, and velocity of
the gas, and the specific total energy $E \equiv |\mbf{v}|^2/2+
p/(\gamma -1)\rho$, with $\gamma= 1.4$.  We assume a time-independent
external potential $\Phi_{\rm ext} \propto v_c^2/(R^2+
a^2)^{1/2}$, where $a=2.5$ kpc is a core radius of the potential and
$v_c = 63$ km s$^{-1}$ is
the maximum rotational velocity, and they are determined to 
mimic the rotation curve of the LMC (\cite{KIM98}).
Since the total gas mass is about 10 \% of the total dynamical mass (see
below), the effect of self-gravity of the gas to the 
rotation curve is not significant. In fact the rotation curve after the system 
evolves is close to the rigid rotation (see \S 3.3 and 
the position-velocity diagrams in Fig. 10).

  We also assume a cooling function
$\Lambda(T_g) $ ($10 < T_g < 10^8$ K)  (\cite{SN}). 
The cooling processes taken into account are, (1) recombination of H, He, C, O, N, Si and Fe, (2) collisional excitation of HI, CI-IV and OI-IV, (3)
hydrogen and helium bremsstrahlung, (4) vibrational and rotational excitation of H$_2$ and (5) atomic and molecular cooling due to 
fine-strucure emission of C, C$^+$ and O, and rotational
 line emission of CO and H$_2$.
As a heating source, $\Gamma_{\rm UV}$,
we assume a uniform UV radiation field  (\cite{GI}), which is
normalized to the local interstellar value, and photoelectric
heating by grains and PAHs. The bulk of the heating for the $10^2-10^4$
K gas is provided by photo-emission of UV irradiated dust grains
 (\cite{BT94}).

We consider two feedback
effects of massive stars on the gas dynamics, namely stellar
winds and supernova explosions, although our results show that 
the former is less
effective than the latter.
 We first identify cells which satisfy
criteria for star formation. The criteria are a surface density
threshold $(\Sigma_g)_{i,j} > \Sigma_c$ and a
critical temperature $(T_g)_{i,j} < T_c$ below which star formation is
allowed. The surface density is defined as $\Sigma_g =2 H \rho$,
where $H$ is the scale height and assumed to be constant (100 pc).
 In these simulations we take 
$\Sigma_c = 40 M_\odot$ pc$^{-2}$ and $T_c = 15$ K (model `Star
Formation 1', hereafter `SF1') and ten times larger threshold density,
$\Sigma_c = 400 M_\odot$ pc$^{-2}$ and $T_c = 15$ K (model SF2).
These criteria are chosen to satisfy the condition, $L_J < 0.1 \Delta$
or $L_J < 0.01 \Delta$, where $L_J$ and $\Delta$ are the Jeans length
and the size of each cell.
The volume filling factors of the star-forming cells to the 
total volume are typically
$\sim 5\times 10^{-3}$ and  $\sim 5\times 10^{-4}$ for model SF1 and
SF2, respectively. Namely the star forming sites are cold, clumpy 
regions. We also assume the star-forming criteria must last during
$10^5$ yr for each cell
before the star formation is initiated.	
Since the spatial resolution is fine enough compared to the GMC size,
we do not have to assume any
global criteria for gravitational instability to identify star forming
sites. Assuming the Salpeter IMF with $m_u = 120 M_\odot$ and $m_l = 0.2
M_\odot$, we create test particles representing massive stars 
($\geq 8 M_\odot$) in the star forming cell.
The kinematics of the test particles in the external potential and 
the self-gravity potential of the gas are traced by the second-order
time-integration method.
 The stars (test particles) inject energy
due to stellar winds during their lifetime which is approximately $\sim
10^7$ yr (\cite{LE98}). 
When one of stars represented by a test particle explodes as
supernova, an energy of $10^{51}$ ergs is injected into the cell as 
thermal energy where the test particle is located at that moment. 
 The cooling procedure is not used for such
cells, but the cells adjoining the supernova cell are treated
normally.  Evolution of the supernova remnant is very dependent on 
its environment. In contrast to past numerical studies of the ISM 
 with supernova explosions on a galaxy scale, we do not introduce
simple evolutionary models for the SNR, such as the Sedov solution,  and 
heating efficiency for the ISM due to a supernova.
With our code, two-dimensional evolution of blast waves caused by
supernovae in an inhomogeneous and 
turbulent medium with global rotation is simulated 
explicitly.  Therefore, we can trace consistently the thermal and
dynamical evolution of the ISM around the star forming regions and the
associated supernovae remnants and superbubbles (c.f. \cite{NI})


The hydrodynamic part of the basic equations is solved by the third-order AUSM
 (Advection Upstream Splitting Method) (\cite{LS}).
After testing this
 code for various hydrodynamical 1-D and 2-D problems, we find that
 AUSM is as powerful a scheme for astrophysical problems as are the
 PPM (\cite{WC}) and Zeus (\cite{SN92}) codes. 
More  details about our numerical code and test results are described in
 Wada \& Norman (2000, in preparation).

We use $1024^2$ Cartesian grid points covering a 8 kpc $\times$ 8 kpc
region. The spatial resolution is 7.8 pc.
A periodic Green function is used to calculate the self-gravity for the
8 kpc $\times$ 8 kpc region with $2048^2$ grid points (\cite{HE}).  The
second-order leap-frog method is used for the time integration.  We
adopt implicit time integration for the cooling term in 
equation (3).

The initial disk is an axisymmetric and rotationally supported
($R=3.7$ kpc) with uniform surface density,
$\Sigma_g = 12 M_\odot $ pc$^{-2}$, 
and the total gas mass is $5\times 10^8 M_\odot$ (\cite{KIM98}).
  Random density and temperature fluctuations are added to the
initial disk. Amplitude of the initial fluctuations 
is less than 5 \% of the unperturbed
values and have an approximately white noise distribution. The initial
temperature is set to $10^4$ K ($R \leq 3.7$ kpc)
 and $10^2$ K ($R > 3.7$ kpc). The reason why we chose the low
temperature in the outer region is to avoid the numerical artifact of 
the boundaries, i.e. reflection and generation of waves or shocks,
 for the initial evolution of the gas disk.
In ghost zones at the boundaries of the 
calculating region (i.e. $8\times 8$ kpc$^2$), all physical quantities remain at their initial
values during the calculations. From test runs we found that this
boundary condition is much better than `outflow' boundaries, because
the latter cause strong unphysical reflection of waves at the
boundaries.

%
\section{RESULTS}
%
\subsection{Evolution and Structure of the ISM}
\subsubsection{Model without Star Formation}

Figure 1 shows the time evolution of density and temperature of
a model without star formation and its feedback 
(hereafter we call this model ``model NSF'').
Due to gravitational and thermal instability in the gas 
disk, clumpy fluctuations evolve in the first $10^8$ yr, and 
then the clumps merge and form larger structure. 
They are deformed by the local tidal field and global shear
 in the non-linear phase,
and as a result, filamentary structure is formed.
Higher density clumps ($ \Sigma_g >10^3 M_\odot$ pc$^{-2}$) 
are formed in the filaments due to the 
gravitational instability, or collisions between the filaments.
The high temperature cavities ($T>10^5$ K) are
formed due to shock heating. 
The dynamical time scale for the large ($L \sim 200$ pc)
 high temperature ($T \sim 10^5$ K)
region seen in the temperature panel at $t=800$ Myr is
$\tau_{\rm dyn} \sim L /\Delta v \sim 200$ pc$/$20 km s$^{-1} \sim 10^7$
yr. The cooling time, $\tau_{\rm cool}$ on the other hand, for this low density region
($n\sim 10^{-3}$) is $\sim 2\times 10^7$ yr (Spitzer 1977),
 which is comparable to 
$\tau_{\rm dyn}$. In fact the high temperature cavity is a tentative
structure that lasts $\sim 10^7$ yr. 
Photo-electric heating and UV irradiation contribute to form $T \sim 10^4$ K gas. 
In the high density ($ \Sigma_g >10^2 M_\odot$ pc$^{-2}$) filaments and 
clouds, the temperature is less than 100 K because the radiative cooling
is effective. 
The global strucure of the disk does not change significantly
after 400 Myr, and 
it reaches a quasi-steady state. 
Figure 2 is time evolution of the volume filling factor, $f_v$, of 
each temperature level and time evolution of 
the maximum density. 
This also represents that the global ($>$ kpc scale) 
system reaches a quasi-stationally state after $t\sim 300$ Myr.
This period is comparable to the local free fall time $t_{ff}$ for
the initial density, $t_{ff} \sim 200 (\Sigma_g/12 M_\odot$
pc$^{-2})^{-1/2}$ Myr.
At the global quasi-steady state, the warm gases ($T_g = 100-10^4$ K) occupy
a large volume ($\sim$ 60 \% of the whole calculating region), and 
$f_v \sim $ 20 \% for the cold ($T_g < 100$ K) and $f_v \sim $ 20 \% for 
hot gas ($T_g = 10^4-10^5$ K).
The maximum density in the same plot 
show slower evolution after $t\sim 300$ Myr than 
the non-linear evolutional phase.

One should note, however, that 
the globally stable state does not mean the local filamentary and porous
structure whose scale is less than 1 kpc.
It is rather quite dynamic and transient, which 
is similar to the past turbulent ISM models
(\cite{BL80}; \cite{CP85}; \cite{CB88}; \cite{RB95}; 
\cite{VZ95a}; \cite{VZ95b}; \cite{GP99}).  
Clump mass is increasing during the initial linear and non-linear
evolutional phase ($t < 200$ Myr) 
due to the mass accretion and merging
with other clumps or filaments, but in the quasi-stationally phase,
disruption processes of the clumps, such as local tidal field (e.g.
interaction between clumps, or shear) or shocks,
prevent monotonic increase of the mass of each clump.
Since we introduce the cutoff temperature for the
cooling (i.e. 10 K), the gaseous pressure prevent
the dense clouds further collapsing towards singularities.
The angular momentum also supports the clouds (see \S 3.3).

The density and temperature structures (Fig. 1) are
 similar to those in the model of Wada \& Norman
(1999), where a smaller disk is investigated (the radius is 1
kpc), but the present model shows kpc-scale inhomogeneity and 
a more asymmetric distribution against the galactic center.
This difference is probably caused by the difference of the rotation
curves used in the two models:
a rigid rotation or differential
rotation. With a rigid rotation, 
random and turbulent motion 
dominates the circular rotation in the central region.
With the global shear, on the other hand, global spirals develop toward 
the galactic center.

\subsubsection{Models with Star Formation}

Figure 3 represents density and temperature maps of the model SF2
at $t= 833$ Myr.  The model SF1, which has ten times smaller threshold
density for the star formation criterion (see \S 2),
 shows very similar density morphology
and temperature structure.
  The red regions in the temperature map are 
hot gaseous regions where $T_g >10^6 $ K. They are young ($<10^6$ yr) supernova
remnants. 
Figure 3 also shows that most young supernova remnants are not axisymmetric. This is caused by that the background ISM is highly inhomogeous, and the radiative cooling at
the dense filaments is so effective, and  that it prevents the blastwaves 
expanding axi-symmetrically. 
Typical size of the hot cavities is less than 500 pc.
Some bubbles together form kpc scale `super bubbles'. The hot region around
 $(-3,-1)$ is one of the examples.  Note that the size of the cavities
is expected to change away from the galactic plane in a 3-D model
(see also \S 4).

Although the filamentary and clumpy structures of the gas 
which result in the star forming models
 are similar to those of model NSF, the kpc scale inhomogeneity seen in the
the model (Fig. 1) is not apparent in the SF models. The large scale
inhomogeneity in the model SF1 and SF2 is less prominent than that of the
model NSF. This is because that the time scale for 
making the kpc-scale holes,
 which  is a dynamical time scale, $\sim 10^8$ yr, is
much longer than the the time scale for supernova explosions
and evolution of the blastwaves, $\lesssim 10^6$ yr.
Approximately $10^3$ supernovae per kpc$^2$ 
explode in this model during $\sim 10^8$ yr.  The kpc-scale
low density cavities, which is formed due to evacuation of gas
by the effect of the gravitational and thermal instabilities,
 cannot evolve under such frequent supernova explosions.

We find that the supernova rate is fluctuate in a time scale
$\sim 10^7$ yr, but it stays in a range
4-10$\times 10^{-4}$ yr$^{-1}$ during 6$\times 10^8$ yr in model SF2
(Fig. 4). This behavior also appears in model SF1, 
but the SN rate becomes smaller for the larger threshold density
($\sim $ 8-12$\times 10^{-4}$ yr$^{-1}$). 
ROSAT observed 46 supernova remnants and more candidates
in the LMC (\cite{HP99}) . If we assume the 
ages of SNRs are between $\sim $ 10000 yrs and $\sim$ 30000 yrs, then 
we have SN explosion every 250 yrs or 750 yrs. This indicates
about 0.0013/yr or 0.004/yr for SN rate in the LMC.
If we use the larger threshold density for the star formation in our
model than that in model SF2, this results in 
smaller supernova rate than $\sim 0.001$ yr$^{-1}$. 
Therefore the threshold density much larger than $400 M_\odot$ pc$^{-2}$
 would be excluded for modeling the LMC, 
if we assume that the star formation
efficiency is 10 \% or less and the standard IMF.


In Figure 5, the star particles at $t=245$ Myr in model SF1 are plotted,
and one may compare it to the density distribution (the right panel)
of the same snapshot. Massive stars are not uniformly distributed, but
they form clusters (`OB associations'). It is notable that the 
distribution of the star clusters does not necessarily correlate to
the inhomogenous gaseous structure. In other words, cavities are
not necessarily associate with the `OB associations'. 
This is also seen in simulations by other groups, for example
2-D simulations of a highly compressible isobaric, non-selfgravitational
 fluid with star formation (\cite{SC99}).

\subsection{Line Emission Maps and Comparison with the Observations}

With the numerical simulations, we have density, temperature, and 
velocity fields (spatial resolution: 7.8 pc).
Using this information as an input, we compute \ion{H}{1} 21 cm brightness map
and CO (J=1-0) line map for model NSF and SF1 and SF2.
 These maps can be directly compared with
the recent high resolution surveys of the LMC: \ion{H}{1} with ATCA (Kim et al.  1998)
and CO (J=1-0) with NANTEN (\cite{FU98}).


With the numerical results described above, the following procedure is
followed to compute a \ion{H}{1} 21 cm brightness map.
It is assumed that atomic hydrogen is in mostly neutral form in those
regions where the temperature is at least a factor of 2 less than 8000 K,
a value typical of (mostly ionized) HII regions, and motivated by measurements
of the kinetic temperatures of HI clouds and HI intercloud gas.
It is also assumed that the \ion{H}{1} level populations of interest
follow a thermal distribution in those regions along the line of
sight, and that this mostly neutral gas dominates the integrated emissivity. 
The line of sight is face-on, i.e., perpendicular to the grid used for the
hydrodynamic simulations. The two-dimensional density indeed is used as a
column density in the radiative transfer calculations. The latter are done
in three dimensions by assuming a scale height for the neutral gas of
H=100 pc, thus converting the column density in a (constant) local density
for each point along the line of sight. It is this three-dimensional grid,
with two-dimensional hydrodynamic information, that is 
used to determine the \ion{H}{1} central brightness temperature in the Monte Carlo
procedure (Spaans 1996), where the ambient radiation field and its interactions
with matter is represented by a discrete number of ``photon packages''.
This method is three-dimensional and explicitly includes optical depth effects
as well as the detailed velocity field of the hydrodynamical simulations 
for photons that travel along lines of sight that are not face-on.
These latter photon trajectories need to be incorporated when the
level populations are not in thermal equilibrium, i.e., for CO.
A local velocity dispersion $\Delta V$ of $1.29\times 10^4 T^{1/2}$ cm s$^{-1}$
is adopted for a kinetic temperature $T$, with a minimum of $0.5$ km s$^{-1}$
due to micro turbulent motions. For each grid point in the hydrodynamical
simulation, the Monte Carlo radiative transfer then yields the integrated
\ion{H}{1} 21 cm intensity along the line of sight.


For the CO (J$=$1-0) line, the approach is similar to the \ion{H}{1} case, with the
appropriate correction factor in $\Delta V$ for the different atomic
weight of the CO molecule.
Furthermore, a carbon chemistry is added to compute the abundance of
CO (\cite{SV97}), for a dust abundance equal to $1/5$ of
Solar. This chemistry is well understood (\cite{VB88}),
and care has been taken to include, for each computed line of sight and
corresponding column density, the important self-shielding transitions of
H$_2$ and CO (c.f. \cite{ST94}). The ambient average interstellar
radiation field, required for the ambient chemical balance, is determined by
scaling with the LMC B band surface brightness in mag per square arcsecond
with respect to Galactic. This typically yields an enhancement of a factor
of $3-5$ in the mean LMC energy density compared to Galactic. It is assumed,
because the hydrodynamical simulations are two-dimensional, that, to compute
the local CO emissivity and the ambient chemical equilibrium, the gas density
is given by the particular line of sight column density divided by a scale
height of $H=100$ pc for the neutral gas. The latter number $H$ does not
strongly influence the qualitative features in the presented maps.

Figure 6 presents \ion{H}{1} 21 cm and CO (J=1-0) line brightness temperature maps 
calculated from the numerical data of the NSF model at $t=800$ Myr, 
using the procedure described above.
Figure 7 is the same plot as Fig. 6, but for model SF2 at $t=834$ Myr.
Size of the \ion{H}{1} filaments, shells, and holes in model NSF is
$\sim$ 1 kpc. The filaments and cavities in the outer region ($R > 2$
kpc) of model SF2 are also kpc-scale.
We would like to emphasize
 that large cavities ($>$ kpc) and filaments in model NSF are NOT caused by a direct dynamical effect 
of star forming activities, but by the non-linear development of
 gravitational
and thermal instabilities.

The CO emission is localized in many clumps (size $\sim$ 10-100 pc) or
clouds complexes (size $\sim$ 0.1-1 kpc) in model NSF.
 Such CO ``cloud'' complexes could be sites for active star forming regions, such as 
the 30 Dor star forming region in the LMC.
Model SF2 shows more uniform distribution of ``stars'' than in model NSF.

Figure 8 (a) and (b) are mass spectra of {\it molecular clouds} obtained from
the CO brightness temperature ($T_B$) distribution  of
models NSF and SF2.
We identify {\it clouds} and their mass by the following procedure.
Assuming a threshold $T_B$ for the CO map (Figs. 6 and 7), then
we have many {\it islands} of CO emission. Most {\it islands} 
are not round in shape, but elongated or filament-like shape.
We derive the mass of the {\it islands} using the surface 
density of the simulation data. Therefore the mass in Fig. 8 is the 
total gas mass of each {\it island},
not the virial mass. Here we plot three histograms for
three different thresholds. 
The mass spectra shows a power-law, which is
roughly $dN_c/dM_c \propto M_c^{-1.7}$ for model NSF, where $dN_c$ is 
number of clouds between the mass $M_c$ and $M_c +dM_c$, but this slope does
not significantly depend
on the threshold brightness. 
On the other hand, the spectrum of 
the star forming model steeper than the model without the energy
feedback especially for small threshold:
$dN_c/dM_c \propto M_c^{-1.7}$, $M_c^{-2.3}$, and 
$M_c^{-2.7}$ for model SF2 with
$T_B=$ 100, 50, and 30 K, respectively.
The behavior of the slope to the threshold $T_B$ in the model NSF
and SF2 means that the stellar energy feedback changes structure of
low density envelope of the dense clumps.

The NANTEN survey discovered about hundred molecular clouds in 
the LMC, and the mass spectrum of them is 
$dN/dM_{\rm vir} \propto M_{\rm vir}^{-1.5\pm0.1}$ for a range
between $10^{5-6} M_\odot$ (\cite{FU98}), where $M_{\rm vir}$ is 
the virial mass of the cloud estimated by using the observed line width.
In our  model, smaller clouds ($M_c < 10^5 M_\odot$) tend to have smaller
mass compared to their virial mass estimated from their internal velocity dispersions. 
In other words, the smaller clouds are not in equilibrium, but
rather in a transient phase (\cite{VZ95a}).
We find a rather weak but positive correlation between the cloud mass $M_c$ and 
the virial mass $M_{\rm vir}^{1/2}$ in our model.
The mass spectrum of the model SF2 would be 
$dN_c/dM_{\rm vir} \propto M_{\rm vir}^{-1.4}$, if we use the lowest
threshold $T_B=30$ K.
A similar discussion for the interpretation of the cloud mass spectra based on 
two-dimensional numerical simulations of the interstellar medium has been given by
V$\acute{\rm{a}}$zquez-Semadeni et al. (1997) (see \S 4).

The maximal mass of the clouds is approximately $10^{6.5} M_\odot$
and $10^{5.5-6} M_\odot$ in models NSF and models with the star formation
(SF1 and SF2).
This implies that very massive clouds ($> 10^6 M_\odot$) are difficult to grow under the frequent 
supernova explosions.
The maximal masses of the observed molecular clouds are
about $10^{6.5} M_\odot$ (\cite{FU98}), which seems to prefer
the model NSF. 
However, this does not mean that the star formation and its energy 
feedback are not important for shaping the ISM 
in the LMC, but it implies that the largest clouds
could evolve in preferentially an environment 
where SNe are not frequent.
Though one should be careful to make comparison between the 
observed and numerical mass spectrum, because the definition and
identification of the ``cloud'' is not exactly identical, and 
it is related to noise level in observations. 

We also find that the \ion{H}{1} map of the computational model shows 
similar statistical properties to the observed \ion{H}{1} map.
Figure 9 shows the \ion{H}{1} luminosity function, i.e. the 
histogram of the observed and numerical \ion{H}{1} map of the LMC.
The \hi\ brightness temperature has been computed from $T_B = {S_\nu
\lambda^2}/{2k_B\Omega_{sb}}$. $S_\nu$ is the \hi\ flux density, $k_B$ is
the Boltzmann constant, and $\Omega_{sb}$ is the solid angle of the 
synthesized beam of ATCA mosaiced map. The \hi\ intensity is determined from 
the integral of the brightness temperatures $\int T_B dv$ over the peak \hi\ 
line profile, where $dv$ is the channel width in kilometers per second.
The \ion{H}{1}
 luminosity functions appear to be log-normal like 
distribution for both observed and model NSF. The distributions of 
the \ion{H}{1} luminosity function of the model NSF and SF2 for lower 
surface brightness ($T_B < 1000$ K)
are similar to the observed one. 
The model SF2, on the other hand, shows excess above $T_B > 1000$ K than
the observations. These high brightness regions are originated from
shock compressed, dense gas. 
\subsection{Position-Velocity Diagrams}
Position-Velocity (PV) diagrams give us information on
the kinematics of the ISM. 
In Fig. 10 (a), we show the PV diagram, in which y-component of
velocity ($v_y$) is integrated through $x$-positions,
for the gas $\Sigma_g < 10^2 M_\odot$ pc$^{-2}$ in model NSF.
Many arc-like structures in Fig. 10 (a) 
look like expanding shells originating from explosional phenomena in the ISM. 
However, they are NOT caused by
explosions because there is no energy input due to supernovae in 
model NSF. 
The arcs in the PV diagram 
are actually caused by the gases that form filament-like structure seen 
in the density map (Fig. 1).
 The filaments and shells exhibit non-circular motion of the order of 
10 km s$^{-1}$, and their shape continuously changed. 
Here we would like to emphasize that, 
 it is hard to distinguish, on the PV diagrams, between
 expanding shells and such shell-like
structures which are changing in shape due to local random motion.

From Fig. 10 (b),  
which is the same as Fig. 10(a), but for 
$\Sigma_g > 10^2 M_\odot$ pc$^{-2}$,
we find that the dense and compact clumps, rotating with
$\sim 10-30$ km s$^{-1}$, which are recognized
 as steep `dotted lines' on the
PV-diagram.  Here we can
identify about 80 clumps, and about half of them show retrograde rotation
against the sense of galactic rotation. A representative one can be seen
at $(x,v_y) = (-0.5, -50)$ and less prominent one is at $(1.5,25)$ and
$(3,75)$. 
Rotation of the clouds is important for the internal structure,
motion, and star formation in the molecular clouds.  
Unfortunately, the spatial resolution in the NANTEN survey ($\sim 40 $
pc) is insufficient to resolve the rotation of each molecular
clouds of the LMC.
High resolution observations and statistical analysis of kinematics of
molecular clouds in external galaxies are necessary to understand
the formation of molecular clouds and star formation processes.

Figure 10 (c) shows the similar features shown in Fig. 10 (a), but for model SF1 at $t=$ 610
Myr. Arc-like structures seen in the PV map for the NSF are not
clearly shown in this map. This means that 
the line-of-sight velocity field is much more random and chaotic
than the NSF model, and 
there is no prominent large-scale coherent motion
in the ISM of the model with stellar energy feedback. 

%
\section{DISCUSSION}
%
The comparison of our model calculations with the observations of the LMC 
indicates that the star formation models are more consistent with the formation
of small ($\ll$ kpc ) and hot bubbles ($T_g$ $>$ 10$^7$ K), which have been detected 
by X-ray diffuse emission. The power law slope of cloud mass spectrum of the
models is close to that of observed one. On the other hand, the model calculations 
without the stellar feedback show even better agreement with global 
inhomogeneity
of the HI and CO gas in 
the LMC, and also
the HI distribution function is well represented in the model NSF.
Therefore, 
it is most likely that
energy feedback to the ISM from massive stars and
supernovae explosions are important processes for producing hot gas, 
but they do not necessarily dominate dynamics and 
formation processes for the large scale inhomogeneity observed 
in \ion{H}{1} and 
CO gas in the LMC.
However,  one should note that the initial condition of the 
models (i.e. axisymmetric and uniform
density distributions) causes nearly uniform star formations
in the whole disk.
If we begin the simulations from
a more inhomogeous disk, stars will be formed non-uniformly, and
effects of the stellar feedback are different,  depending on location of
the disk. 
We suspect that the more realistic model of
the ISM in the LMC-type galaxy would be
between our two extreme cases, but that can not be achieved by
changing the threshold density of the star forming model.

In our models with stellar feedback, 
typical sizes of the hot cavities are less than 500 pc.
Some `superbubbles' can be 
formed in a outer disk region, and their
linear sizes are $\sim$ kpc (see Fig. 3).
Rosen \& Bregman (1995) revealed in their two-dimensional, two-fluid (stars
and gas) simulations in a disk galaxy that hot bubbles are formed by
stellar activities, and their sizes depend on the energy injection rate.
In their simulations, the linear size of the 
bubbles is up to $\sim 500-1000$ pc for the supernova rate is 0.0075-0.03 
yr$^{-1}$. This rate is 1-2 orders of magnitude larger than that of
our star forming models.
Local, three dimensional MHD simulations of the
ISM with supernova explosions by 
Korpi et al. (1999) show that the linear size of the superbubbles are 
typically 200-400 pc, which is consistent with our SF models.
Gazol-Patino \& Passot (1999) compute the evolution of the ISM 
in a region of the Galactic plane of size 1 kpc$^2$ in two-dimensional periodic 
domain. They found that superbubbles, and the largest one has linear scale 
is $\sim$ 500-1000 pc, due to about 3000 supernova explosions
in $\sim 10^7$ yr.  The background density of their simulation
is equivalent to that in the outer region of our simulations.
The local ($<$ kpc scale) structure of the ISM of our
global simulations, where hot bubbles and cold filaments coexist,
is also similar to the local, 2-D simulations of the ISM with 
a periodic boundary condition (\cite{VZ95a}). 
In a conclusion, the past local models with supernova 
explosions, in which superbubble formation is reported, consistent with
our global models with stellar energy feedback concerning the size of the
superbubbles.

The cloud mass spectra of the model SF1 and SF2 are 
steeper than that of the model NSF (Fig. 8).
In other words, the stellar energy input affects the evolution of 
molecular clouds.
V$\acute{\rm{a}}$zquez-Semadeni, Balleseros-Paredes, \& Rodriguez (1997) (hereafter VBR97) analyzed their 
2-D hydromagnetic simulations and showed that the mass
spectra have the form $dN_c/dM_c \propto M_c^{-1.44\pm0.1}$
(c.f. their Fig. 3 and our Fig. 8), and they reported that the mass
spectra from observations are 
consistent with that from simulations in which the
density filed had been shaped by stellar activity.
Note that our mass spectrum of the star formation models, the slope is 
shallower for smaller clouds ($ M_c < 10^5 M_\odot$) than that for 
massive clouds ($dN_c/dM \propto M^{-2}$). 
Therefore our results might be consistent with the results of
VBR97 (see also \S 3.2).
However, direct comparison between these two results
 should not be straightforward,
because there are number of differences between our simulations and
theirs, on the numerical scheme (AUSM vs. pseudo spectral method), the 
boundary condition (global simulations vs. local periodic boundary),
and the cooling curve especially for $T_g < 100$ K ($\Lambda \ne 0$
vs. $\Lambda = 0$).
The maximum density contrast is about 5000 in their model, but 
it is about $2\times 10^6$ in our model.
This difference is caused probably due to the difference of
 the cooling curve for
the cold gas, and also due to the numerical scheme.
In their numerical method, they added a mass
diffusion term to the continuity equation in order to smooth out 
the density gradients (see also \cite{VZ95a}).
This could affect the structure of shocks and dense gas, i.e, the cloud mass
spectrum. 
The recipe for the stellar energy feedback are also different.
VBR97 assumed that once a star is formed, then it remains 
fixed with respect to the numerical grid. In our models, the star
particles are orbited in the self-gravitational potential of the gas and
the external fixed potential.
The last point, however, would not be important, if number of stars is large
enough, or the ISM is fully turbulent.
VBR97, on the other hand, include the magnetic field in their models.
Therefore, again, one should be careful to make direct comparison between 
results in the present paper and VBR97.

Our result implies that there are two mechanisms of cavity formation.
One is the pure evacuation of gas from the low-density regions by the
effect of the gravitational and thermal instabilities. Similar 
evacuation phenomena (\cite{EG94}; \cite{VZ96}) are also observed in the simulations of the turbulent ISM 
without SNe (\cite{VZ95b}). The other is the formation of cavities by
the effect of SNe, and especially by the synchronized explosions of
stars in OB associations. The two processes are essential to produce
the whole structure of the ISM. The latter is necessary for hot ($T_g > 10^6$ K) 
component of the ISM. These two type of low-density regions are also observed
in the two-dimensional, local simulations of the turbulent ISM with SNe 
by Gazol \& Passot (1999).

As mentioned in \S 1, the origin of supergiant \ion{H}{1} holes in 
galaxies has been controversial (see also \cite{WB99}).
If the observed kpc-scale holes and shells are caused by explosional
phenomena only, one needs 
highly energetic events, like Gamma Ray Bursts (\cite{LP99}).
However, our numerical simulations give an alternative explanation
about the origin of the large scale holes: 
the non-linear evolution of the multi-phase gas disk.  
In the multi-phase ISM, most of the gas mass is concentrated
in the cold, dense clumps and filaments, but volume filling 
factor of such component is much smaller than that of
hot, diffuse gas. 
The hot regions are surrounded by dense filaments as seen in
Fig. 1 and Fig. 3.
Therefore the multi-phase ISM in a quasi-steady state is 
naturally porous. Our results suggest that if the energy feedback
from massive stars are not effective, 
kpc-scale inhomogeneity can be evolved in a disk in about several $10^8$ yr.
Dense filaments are changing in shape due to 
local random velocity field as well as the global shear, and often 
show kinematics features similar to those of ``expanding shells'' as seen in
the position-velocity map (\S 3.3).
We suspect that many supergiant holes and shells 
in dwarf galaxies do not have an explosion origin, but
that supergiant shells far outside the disk plane
can not be formed without explosional events.

The work presented here yields the model for the ISM in a LMC-type galaxy. 
Nevertheless there are couple of things that one should consider for constructing  
a complete numerical model of the LMC including the interaction with 
the Small Magellanic Cloud,
the optical bar, and non-uniform UV radiation field due to 
active star forming regions, such as the 30 Dor regions. 
The interactions with the SMC 
are important events for the star formation history 
in the LMC and the Magellanic Stream (\cite{MF80}; \cite{VA96}; \cite{GN96}).
The off-center optical bar could also perturb the global structure of
the ISM and star formation in the LMC (\cite{GA98}). However the 
\ion{H}{1} mapping (\cite{KIM98}) does not show clear evidence of the 
stellar bar, namely off-set shocks which are often seen in barred galaxies.
It would be interesting to investigate the effect of the 
off-center stellar bar in our model.  
This might contribute to the formation of the active star forming
region, such as 30 Dor region.

In the present paper, we have not solved the vertical structure
of the ISM. 
It is expected that the hot component behaves differently in
three dimensions. The hot gas, which is above $10^6$ K, cannot
be confined in the disk plane (\cite{RB95}; \cite{DV99}),
and it would probably been blown out
from the disk plane. 
The scale-height of the 
hot gas should be larger than the cold gas, 
and, as a result, the volume filling factor of 
the hot gas would be
larger away from the disk plane.
Thus the radiative cooling would be less effective 
in the hot gas, in three dimensions.
This may affect interaction between cold and hot components, 
and the feedback process on the ISM. For example,
it is more difficult to form the supergiant holes by SNe.
In a subsequent paper, we will extend our method to
three-dimensional modeling, and investigate these problems.
%
\section{CONCLUSIONS}
%
Using high resolution hydrodynamical simulations, we have
computed that the global dynamics and structure of the multi-phase
ISM in a LMC-type galaxy.
Due to gravitational and thermal instability in the gas 
disk, clumpy fluctuations evolve, and 
then the clumps merge and form filamentary structure in the non-linear phase.
Higher density clumps are formed in the filaments due to the 
gravitational instability, or collisions between the filaments.
Our numerical model with the star formation allows us to provide
the evolution of the blastwaves due to supernovae explosions in 
the rotating, inhomogeneous, multi-phase, and turbulent-like media.
We find that the supernova rate in the model with stellar energy
feedback is typically of the order of
0.001 yr$^{-1}$ during several hundred Myrs, but fluctuates rapidly
(time scale $\sim$ 10 Myr)  by a factor of three or four.
The model also shows that kpc-scale low density cavities seen
in the observed \ion{H}{1} map (\cite{KIM98}) are difficult to be formed under 
frequent supernovae, but SNe are necessary to form 
hot bubbles where the gaseous temperature is
greater than $10^{6-7}$ K. 
Our result suggests that there are two possible causes of 
low density regions in the ISM.
The kpc-scale inhomogeneity and arcs can be formed
as a natural consequence of non-linear evolution of the multi-phase interstellar medium
in a LMC-type galaxy. We find in the PV-diagram of the numerical models 
that filamentary structure in the non-star forming model, which
are caused by the gravitational instability, has similar kinematics
to the structure formed by SN explosions. The dense clouds are rotating,
and about half of of them show retrograde rotation
against the sense of galactic rotation.
Using the Monte Carlo radiative transfer code, 
we have computed \ion{H}{1} and CO brightness
temperature distributions, and compared them with those from the 
recent observations.
The CO cloud mass spectrum in the model with stellar energy feedback is
similar to the observed one (\cite{FU98}), but 
\ion{H}{1} distribution function is well fitted by the model without 
SNe. 
Therefore we conclude that the small scale structure and dynamics of the ISM
   in the LMC is mainly affected by the stellar activities, but the 
   gravitational instability significantly contributes to the global morphology
   and dynamics of the interstellar matter in a kpc-scale.

\acknowledgments

 We are grateful to Colin Norman for stimulating discussions.
We also thank E. V$\acute{\rm{a}}$zquez-Semadeni (the referee) 
for his fruitful comments and suggestions.
Numerical computations were carried out on VPP300/16R at the
Astronomical Data Analysis Center of the National Astronomical
Observatory, Japan, VPP700 at the SUBARU observatory, Hawaii, and
VPP500 at RIKEN, Japan.
We would like to thank R. Ogasawara for his help at the SUBARU observatory .
KW is supported in part by 
Grant-in-Aids for Scientific Research (no. 1113421)
of Japanese the Ministry of Education, Culture, Sports and Science,
and Foundation for Promotion of Astronomy, Japan.
MS is supported by Hubble Fellowship grant HF-01101.01-97A, awarded by
STScI.

\newpage
\figcaption{Evolution of density and temperature distributions of a
model without star formation and energy feedback (model ``NSF''). 
The density and temperature are Log-scaled, and their units 
are $M_\odot$ pc$^{-2}$ and K, respectively.
Time is shown at each panel in an unit, $10^8$ yr.
Note: Although the gray scale is 
assigned for $1000-0.01 M_\odot$ pc$^{-2}$ in this plot, the maximum density is
much grater than 1000 $M_\odot$ pc$^{-2}$ (see Fig. 2).}

\figcaption{Evolution of the volume filling factor for 
four different phases, i.e. $T \leq 100$ K (filled circles), $100<T \leq 9000$ K
(open
circles), 
$9000<T < 10^5$ K (open diamonds), and $T \geq 10^5$ K (filled diamonds).
The maximum density of the gas is also plotted (thin solid line).}

\figcaption{Same as Fig. 1, but for 
a model with star formation and energy feedback (model ``SF2'') at
$t=834$ Myr. }

\figcaption{Evolution of a total supernova rate in model SF2.}

\figcaption{Star particles and density distribution at $t=245$ Myr in 
model SF1.}

\figcaption{Brightness temperature ($T_B$ (K)) maps of 
\ion{H}{1} (left) and CO (J=1-0) (right) computed from data of model
NSF at $t=800$ Myr. The intensity of \ion{H}{1} is Log-scaled. }

\figcaption{Same as Fig. 6, but for model SF2 at $t=834$ Myr}

\figcaption{(a) Mass spectrum of CO clouds for model NSF for three
thresholds of the brightness temperature, $T_B = 30$ (solid line), 50 (dashed line), and 100 K (dotted line). 
The thick line shows $dN_c/dM_c \propto M^{-1.7}$.
(b) Same as (a), but for model SF2.}

\figcaption{Probability distribution function (pdf) of \ion{H}{1}. Thick
line shows pdf of model NSF ($N (T_B)$ is the number of cells for the
brightness temperature, $T_B$, and $N_0$ is the total number of cells).
The stars show the histogram from the \ion{H}{1}
synthesis observation, with N$^{1/2}$ error bars, which is normalized
for $\log(T_B) =2.8$ K of the model pdf.}

\figcaption{Position-Velocity maps for (a) low density gas in model NSF, 
(b) high density gas in model NSF, and  (c) low density gas in model SF1.}


\end{document}